# Using Dimensional Analysis to Construct Multiple-Scalar-Vector-Tensor Cosmological Field Theories

by

Gregory W. Horndeski
2814 Calle Dulcinea
Santa Fe, NM 87505-6425

e-mail:
horndeskimath@gmail.com

March 28, 2018

# Abstract

In this paper I shall consider various possible scalar-vector-tensor field theories which might be used to describe the Universe. After imposing numerous constraints of a physical and mathematical nature on the theories under consideration, we shall see that there still exist a plethora of viable theories from which to choose. But hopefully I shall be able to convince you that a scalar-vector-tensor field theory involving five scalar fields provides a promising cosmological model. The advantage of this theory is that there is no energy-momentum tensor for the matter fields, nor arbitrary functions, only dimensionless constants, that need to be determined. I shall also construct the most general multiple-scalar-vector-tensor field theory which is dimensionally consistent and free of dimensioned constants. It turns out that this field theory yields second-order field equations.

**Section 1: Introduction**

Einstein once remarked, "what really interests me is whether God had any choice in the creation of the world." Well, if by choice here we mean choice of field equations, then the answer would be No, *if* God limited the category of admissible theories to be those (metric) tensor theories of second-order, derivable from a variational principle. For Lovelock [1] has shown that the only (vacuum) field equations of that type are

$$G^{ij} + \Lambda g^{ij} = 0 , \qquad \text{Eq.1.1}$$

where $G^{ij}$ is the Einstein tensor, and $\Lambda$ is the cosmological constant. A Lagrangian that yields these field equations is

$$L_E := g^{1/2} R - 2 g^{1/2} \Lambda , \qquad \text{Eq.1.2}$$

where if $L = L(g_{ij}; g_{ij,k}; \ldots )$ is a Lagrange scalar density, then its associated Euler-Lagrange tensor density is

$$E^{ij}(L) := -\frac{\partial L}{\partial g_{ij}} + \frac{d}{dx^h}\frac{\partial L}{\partial g_{ij,h}} + \ldots \qquad \text{Eq.1.3}$$

My notational conventions are the same as those employed in Lovelock and Rund [2]. I shall also use geometrized units, in terms of which $c = G = 1$, and dimensioned quantities are measured in terms of length. In terms of these units $\Lambda$ has units of (length)$^{-2}$. Lastly, it will be assumed that we are working in a four-dimensional pseudo-Riemannian space with Lorentzian signature (-+++).

For many years Eq.1.1 formed the basis of Einstein's theory of gravity (with or without the cosmological constant Λ). But as the years passed since Einstein introduced his theory, some physicists came to realize that we require more fields then just a tensor field to describe gravitational effects in the U (:=Universe). To that end in 1961 Brans and Dicke [3], introduced a ST (:=scalar-tensor) theory based on the Lagrangian

$$L_{BD} := g^{1/2}(\varphi R - \omega \varphi^{-1} X + \mathcal{L}_M) \qquad \text{Eq.1.4}$$

where $\omega$ is a dimensionless constant, $X := g^{ab}\varphi_{,a}\varphi_{,b}$ and $\mathcal{L}_M$ is the matter Lagrangian, which, following custom, has units of (length)$^{-2}$. (The BD (:=Brans-Dicke) theory was built upon an earlier ST theory by Jordan [4], which was not the first ST gravitational theory.) In BD theory $\varphi^{-1}$ is interpreted as the locally measure gravitational constant.

In the early 1970's my Ph.D. supervisor, Professor Lovelock, and I addressed the problem of the uniqueness of second-order ST field theories derivable from a variational principle. This investigation lead to two papers, Horndeski and Lovelock [5], and Horndeski [6]. The latter paper was drawn from my thesis, and demonstrated that ∃ a cornucopia of second-order ST field theories available from which to choose. At that time this observation led Lovelock and me to believe that ST theories probably will be of no value in describing the U. Our thinking was that clearly anything as unique as our U would require unique equations to describe it, and ST theories were anything but. So the results I presented in [6] "remained a sort of theoretical curiosity for more than thirty years" (in the words of Bettoni and Liberatti [7]), before being rediscovered within the context of Galileon theory (which was introduced by Nicolis, *et.al.* in [8]). [6] became the basis of a great deal of research into what became known as Horndeski Scalar Theory (=:HST). The hope was that the scalar field could be used to study such things as inflation and dark energy, which were phenomenon that did not even exist when Lovelock and I were doing our research into ST theory. The Lagrangian $L_H$ that forms the basis of HST can be written as the sum of four Lagrangians

$$L_H = L_2 + L_3 + L_4 + L_5$$

called the quadratic, cubic, quartic and quintic Horndeski Lagrangians. Each Lagrangian has a scalar coefficient function of φ and X, denoted by $G_.(\varphi,X)$. (*See Horndeski Theory in Wikipedia for the explicit form of $L_H$.*)

The recent observation of a pair of colliding neutron stars on August 17, 2017, showed that the speed of gravitational waves, $c_g$, must equal 1, to 1 part in $10^{15}$. Baker, *et.al.* [9], Creminelli and Vernizzi [10], Sakstein and Jain [11], and Ezquiaga and Zumalaćarregui [12], have analyzed what effect this observation has on the $G_.$'s appearing in $L_H$, and have shown that for all practical purposes $G_5$ must vanish, while

$G_4$ must be independent of X. So under these restrictions $L_H$ reduces to

$$L_{rH} := g^{\frac{1}{2}}[G_2(\varphi,X) + G_3(\varphi,X)\Box\varphi + G_4(\varphi)R] .\qquad \text{Eq.1.5}$$

If $G_4$ is a constant in Eq.1.5, then, loosely speaking, $L_{rH}$ reduces to a Lagrangian of the Einstein form with a scalar field; *viz.*,

$$L_{ErH} := g^{\frac{1}{2}}[\kappa R + G_2(\varphi,X) + G_3(\varphi,X)\Box\varphi],\qquad \text{Eq.1.6}$$

where $\kappa$ is a constant, which I take to be unitless, since I prefer to not introduce any new dimensioned constants. However, if

$$\frac{dG_4}{d\varphi} \neq 0,$$

then the scalar field in Eq.1.5 can be "renormalized" so that $G_4 = \varphi$. This yields a generalized Brans-Dicke type Lagrangian

$$L_{BDrH} := g^{\frac{1}{2}}[\varphi R + G_2(\varphi,X) + G_3(\varphi,X)\Box\varphi] .\qquad \text{Eq.1.7}$$

The Lagrangians $L_{ErH}$ and $L_{BDrH}$ still have a lot of arbitrariness in them which makes them valuable to people working in cosmology. However, that arbitrariness is what makes them unsavory for those of us who covet a (somewhat) unique set of equations governing the U. In the next section I shall show how techniques from dimensional analysis, developed by Adersley in [13], can be used to determine the functional form of $G_2$ and $G_3$. This investigation will naturally lead me to introduce four more scalar fields, and also a vector field, in an attempt to find suitable cosmological field equations for the U. In the third section I shall derive the form of all multiple-scalar, (single) vector-tensor field theories which are dimensionally consistent, flat space compatible (to be defined later), and free of dimensioned constants. It will be seen that all such theories yield second-order field equations. The paper will conclude with some remarks on what I have accomplished here, and their possible implications and extensions.

**Section 2: Dimensional Analysis of Scalar-Tensor Field Theories**

I shall begin this section with a dimensional analysis of the Lagrangians presented in Eqs.1.6 and 1.7. Recall that in geometrized units it is customary to assume that $g_{ab}$ and $\varphi$ are unitless, and that the coordinates are chosen to have units of length. If a quantity $\Phi$ has units $l^\alpha$ ($l :=$ length) we shall write $\Phi \sim l^\alpha$ (read: $\Phi$ has units of $l^\alpha$). If $l'$ is another length scale then there exists a positive real number $\lambda$ which is such that $l' = \lambda l$. So if $\Phi'$ denotes the value of $\Phi$ in terms of the $l'$ units, then we have $\Phi' = \lambda^\alpha \Phi$. Now suppose that $\Phi$ is a concomitant of quantities A, B, ..., where $A \sim l^a$, $B \sim l^b$, .... The ADA (:=Axiom of Dimensional Analysis) requires that for every $\lambda > 0$

$$\Phi' = \Phi(A', B', \ldots) = \lambda^\alpha \Phi(A, B, \ldots),$$

and so

$$\Phi(\lambda^a A, \lambda^b B, \ldots) = \lambda^\alpha \Phi(A, B, \ldots). \qquad \text{Eq.2.1}$$

Eq.2.1 is a powerful restriction on the form of the concomitant $\Phi$, and, in many cases, it determines the functional form of $\Phi$. It must be noted that when doing dimensional analysis, if there are dimensioned constants involved in determining the value of $\Phi$, they have to be stipulated as arguments of $\Phi$. We shall now require that $G_2 = G_2(\varphi,X)$ and $G_3 = G_3(\varphi,X)$ satisfy the ADA, and examine the implications of doing so.

Since $g^{\frac{1}{2}}R \sim l^{-2}$, and $g^{\frac{1}{2}}\varphi R \sim l^{-2}$ in Eqs. 1.6 and 1.7 respectively, we can conclude that $G_2 \sim l^{-2}$ and $G_3 \sim l^0$ in Eqs. 1.6 and 1.7. By assumption $\varphi \sim l^0$ and $g_{ab} \sim l^0$, and hence $X \sim l^{-2}$. Consequently Eq.2.1 implies that $G_2$ and $G_3$ must satisfy

$$\lambda^{-2} G_2(\varphi,X) = G_2(\varphi, \lambda^{-2}X) \quad \text{and} \quad G_3(\varphi,X) = G_3(\varphi, \lambda^{-2}X) . \qquad \text{Eq.2.2}$$

Upon differentiating Eq.2.2 with respect to $\lambda$, and then setting $\lambda=1$, we obtain two simple partial differential equations which can be solved to show that

$$G_2 = \xi(\varphi)X \quad \text{and} \quad G_3 = \zeta(\varphi)$$

where $\xi$ and $\zeta$ are arbitrary differentiable functions of $\varphi$. Consequently if we require $L_{\text{ErH}}$ and $L_{\text{BDrH}}$ to satisfy the ADA they must be given by

$$L_{\text{ErH0}} := g^{\frac{1}{2}}[\kappa R + \xi(\varphi)X + \zeta(\varphi)\Box\varphi] \qquad \text{Eq.2.3}$$

and

$$L_{\text{BDrH0}} := g^{\frac{1}{2}}[\varphi R + \xi(\varphi)X + \zeta(\varphi)\Box\varphi] , \qquad \text{Eq.2.4}$$

where the subscript 0 on the above Lagrangians is there to indicate that we are assuming that $\varphi \sim l^0$. The $g^{\frac{1}{2}}\zeta\Box\varphi$ terms in Eqs.2.3 and 2.4 can be expressed as a divergence plus a term of the form $g^{\frac{1}{2}}\xi X$. Hence Eqs.2.3 and 2.4 can be equivalently

expressed as

$$L_{ErH0} = g^{1/2}[\kappa R + \eta(\varphi)X] \qquad \text{Eq.2.5}$$

and

$$L_{BDrH0} = g^{1/2}[\varphi R + \eta(\varphi)X], \qquad \text{Eq.2.6}$$

where $\eta$ is an arbitrary function of $\varphi$. Eq.2.6 is just a slight generalization of the Brans-Dicke Lagrangian, while Eq.2.5 is the usual Lagrangian employed for a scalar field in Einstein's theory, after renormalizing $\varphi$. So we see that we really have not gotten anything new by applying the ADA when $\varphi \sim l^0$, which agrees precisely with a special case of Aldersley's results, found in [13].

It is interesting to note what happens when we apply dimensional analysis techniques to the quartic and quintic Lagrangians of Horndeski theory. Both $L_4$ and $L_5$ have units of $l^{-2}$. This implies that $G_4 \sim l^0$ and $G_5 \sim l^2$. Assuming that $L_4$ and $L_5$ satisfy the ADA, it is easy to demonstrate that

$$G_4 = G_4(\varphi) \text{ and } G_5 = \mu(\varphi)X^{-1},$$

where $\mu$ is an arbitrary function of $\varphi$. If you want $L_5$ to be well defined for all choices of $\varphi$, including $\varphi$ being null, then you would have to take $\mu = 0$, in which case $L_5$ vanishes. Thus dimensional analysis applied to $L_4$ and $L_5$ leads to the same restrictions on $G_4$ and $G_5$ as those obtained in [9]-[12], where they were looking for conditions on the coefficients of $L_H$ that lead to $c_g = 1$. However, due to the above work we know that the demand that $L_H$ satisfies the ADA is much more severe than the demand that it leads to $c_g = 1$. This follows from the fact that the ADA restricts the form of $G_2$ and $G_3$, while the requirement that $c_g = 1$, does not. Nevertheless, one can't help but wonder whether any field theory that involves the combination of unitless tensor fields with the metric tensor, and which satisfies the ADA, will automatically lead to $c_g = 1$.

Let's now try a different approach in our investigation of the Lagrangians $L_{ErH}$ and $L_{BDrH}$. Instead of viewing $\varphi$ as a possible additional gravitational field, let us identify it with the locally measured dark energy determined by a fixed set of observers permeating U. (These observers generate a 3 + 1 decomposition of U.) In that case $\varphi \sim l^1$, $X \sim l^0$, $L_{ErH} \sim l^{-2}$ and $L_{BDrH} \sim l^{-1}$. If we now apply the ADA to these two Lagrangians using the technique described above, we find that when $\varphi \sim l^1$, $L_{ErH}$ and $L_{BDrH}$ become

$$L_{ErH1} := g^{\frac{1}{2}}[\kappa R + \xi(X)\varphi^{-2} + \zeta(X)\varphi^{-1}\Box\varphi],  \qquad \text{Eq.2.7}$$

and

$$L_{BDrH1} := g^{\frac{1}{2}}[\varphi R + \xi(X)\varphi^{-1} + \zeta(X)\Box\varphi] \qquad \text{Eq.2.8}$$

where $\xi$ and $\zeta$ are arbitrary functions of X. The subscript 1 on the above Lagrangians is there to remind us that we are assuming that $\varphi \sim l^1$. It should be noted that in general the terms involving $g^{\frac{1}{2}}\Box\varphi$ in Eqs.2.7 and 2.8, can not be absorbed into a divergence plus first order terms, as was the case when $\varphi \sim l^0$.

Recall that when we began our investigation of second-order scalar-tensor field theories we had four Lagrangians, each with an arbitrary function of φ and X. Now we have two Lagrangians, each with two arbitrary functions of X. At this point the conventional thing to do to get a Lagrangian which might be able to describe the U is to add $\mathcal{L}_M$ (the matter Lagrangian) to $L_{ErH1}$ and $L_{BDrH1}$ and see if we can choose values for $\xi(X)$ and $\zeta(X)$ which yield viable theories. Rather than take that approach let us stop and consider what fields we really need to describe the U cosmologically. At any event in spacetime we are going to encounter dark matter, as well as ordinary matter, neutrinos and photons, and possibly electromagnetic fields. Typically these quantities appear in the energy-momentum tensor derivable from $\mathcal{L}_M$ (*see, e.g.,* Barreira, *et.al.,*[14] where they introduce a matter energy-momentum tensor that involves the energy densities of the ordinary and dark matter, as well as the energy density of (massive) neutrinos and photons). Rather than do that, I suggest that we completely discard $\mathcal{L}_M$, and instead introduce four scalar fields $\varphi_D$, $\varphi_M$, $\varphi_N$ and $\varphi_P$ which we shall interpret as representing the total energy of dark matter, ordinary matter, neutrinos and photons, respectively, with respect to a fixed class of observers, which arise from a 3+1 decomposition of U. In addition let us include a vector field, $A_a$, to represent electromagnetic fields. Now it would be folly to try to determine the most general Lagrangian which is a concomitant of five scalar fields, a vector field and a metric tensor, and which yields second-order field equations. But what if we looked for such theories that also satisfied the ADA? Unfortunately Aldersley's dimensional analysis techniques fail when $\varphi \sim l^1$ and the other fields $\sim l^0$. That is why in the next section I shall assume that all fields are unitless. So for the present let us assume that the five scalar fields have units of $l^1$ and forget about the vector field for a moment. Let's return to Eqs. 2.7 and 2.8, and use them to try to figure out how to put five scalar fields into the Lagrangian.

The Lagrangian presented in Eq.2.8 has three parts. The obvious way to generalize the first part; *viz.*, $g^{\frac{1}{2}}\varphi R$, is to replace it by

$$g^{\frac{1}{2}}(\sum_{\alpha}\varphi_{\alpha})R, \qquad \text{Eq.2.9}$$

where $\{\varphi_{\alpha}\} \equiv \{\varphi_0, \varphi_1, \varphi_2, \varphi_3, \varphi_4\} := \{\varphi, \varphi_D, \varphi_M, \varphi_N, \varphi_P\}$. Note that in Eq.2.9 the sum over the $\varphi_{\alpha}$'s is the total energy at an event in spacetime, excluding any energy in electromagnetic fields present at that event. The third part of the Lagrangian $L_{BDrH1}$ is $g^{\frac{1}{2}}\zeta(X)\Box\varphi$. If we try to replace this Lagrangian by terms of the form $g^{\frac{1}{2}}\zeta(X_{\alpha})\Box\varphi_{\beta}$, where $X_{\alpha} := g^{ij}\varphi_{\alpha,i}\varphi_{\alpha,j}$ (no sum over repeated Greek indices), then the resulting Lagrangian yields third-order field equations when $\alpha \neq \beta$. This suggests that we replace the third part of $L_{BDrH1}$ by

$$g^{\frac{1}{2}}\sum_{\alpha}\zeta_{\alpha}(X_{\alpha})\Box\varphi_{\alpha}, \qquad \text{Eq.2.10}$$

where the $\zeta_{\alpha}$'s are arbitrary functions of $X_{\alpha}$.

So far things have been going quite well in our modification of $L_{BDrH1}$ to obtain a Lagrangian built from five scalar fields that yields second-order field equations. Now for the second term in $L_{BDrH1}$ which is $g^{½}\xi(X)\varphi^{-1}$. If we just replace this term by a sum over α of terms of the form $\xi_\alpha(X_\alpha)\varphi_\alpha^{-1}$ then the Lagrangian which we would obtain using Eqs.2.9 and 2.10 would involve absolutely no interaction between the five scalar fields. That will never do. We can replace the Xs in ξ by $X_{\alpha\beta} := g^{ij}\varphi_{\alpha,i}\varphi_{\beta,j}$ (note: $X_\alpha \equiv X_{\alpha\alpha}$), but then what do we replace $\varphi^{-1}$ by? The most "natural choice" would be $(\varphi_\alpha\varphi_\beta)^{-½}$, where we really do not need to worry about differentiability problems since we shall assume that $\varphi_\alpha\varphi_\beta > 0$ (or that $X_{\alpha\beta}/\varphi_\alpha\varphi_\beta$ is well behaved). These observations suggest that a good generalization of $L_{BDrH1}$ to the case where there exist five scalar fields would be

$$L_{BD5ST} := g^{½}[\sum_\alpha \varphi_\alpha R + \sum_\alpha \zeta_\alpha(X_\alpha)\Box\varphi_\alpha + \sum_{\alpha \leq \beta} \xi_{\alpha\beta}(X_{\alpha\beta})(\varphi_\alpha\varphi_\beta)^{-½}] , \qquad \text{Eq.2.11}$$

where $\zeta_\alpha(X_\alpha)$ and $\xi_{\alpha\beta}(X_{\alpha\beta})$ are arbitrary differentiable functions of the indicated arguments (so, *e.g.* $\xi_{12}(X_{12})$ is just a function of $X_{12}$, and not a function of all 15 $X_{\alpha\beta}$). Now, of course, you could come up with other generalizations of $L_{BDrH1}$. *E.g.,* we could replace the coefficient of R in Eq.2.11, by $[\sum_\alpha \varphi_\alpha]^r / [\varphi_0\varphi_1\varphi_2\varphi_3\varphi_4]^{(r-1)/5}$, where r is any positive real number. But what I have tried to do was generalize $L_{BDrH1}$ in a way that was more akin to its original form.

Using arguments similar to those just employed, a "natural" generalization of $L_{ErH1}$ is given by

$$L_{E5ST} := g^{1/2}[R + \sum_\alpha \zeta_\alpha(X_\alpha)\varphi_\alpha^{-1}\Box\varphi_\alpha + \sum_{\alpha\leq\beta} \xi_{\alpha\beta}(X_{\alpha\beta})\varphi_\alpha^{-1}\varphi_\beta^{-1}] \,, \qquad \text{Eq.2.12}$$

where, as above, $\zeta_\alpha(X_\alpha)$ and $\xi_{\alpha\beta}(X_{\alpha\beta})$ are differentiable functions of the indicated arguments.

The simplest choice for the coefficient functions $\zeta_\alpha$ and $\xi_{\alpha\beta}$ would be $\zeta_\alpha = k_\alpha X_\alpha$ and $\xi_{\alpha\beta} = k_{\alpha\beta} X_{\alpha\beta}$, where the k's are unitless constants, some of which can probably be set equal to zero. For this simple choice of coefficients, we see that God still had quite a bit of choice when it came to choosing suitable field equations. But at least all God had to do was choose from numerical constants. This was probably not the answer that Einstein was looking for.

Let's now turn our attention to the problem of adjoining external electromagnetic fields to the above Lagrangians. I investigate this problem for the case of a (single) scalar-tensor field theory in [15]. Since the usual Lagrangian employed to derive Maxwell equations; *viz.,*

$$L_M := -g^{1/2} F^{ab}F_{ab} \,, \qquad \text{Eq.2.13}$$

($F_{ab} := A_{a,b} - A_{b,a}$) is conformally invariant, in [15] I investigated the problem of constructing all conformally invariant scalar-vector-tensor field theories that were consistent with conservation of charge and flat space compatible. The upshot of that analysis was that the most general vector field equations satisfying those conditions could be obtained from the following Lagrangian

$$L_{GM} := g^{\frac{1}{2}}\beta(\varphi)F^{ab}F_{ab} + \gamma(\varphi)\varepsilon^{abcd}F_{ab}F_{cd} \qquad \text{Eq.2.14}$$

where $\beta$ and $\gamma$ are arbitrary functions of $\varphi$, and $\varepsilon^{abcd}$ is the Levi-Civita tensor density. The corresponding vector field equations are

$$E^a(L_{GM}) = 4g^{\frac{1}{2}}[\beta F^{ab}{}_{|b} + \beta'F^{ab}\varphi_b] + 4\gamma'\varepsilon^{abcd}\varphi_b F_{cd} \qquad \text{Eq.2.15}$$

where a prime denotes a derivative with respect to $\varphi$.

Now we wish to modify $L_{GM}$ for the case where there are five scalar fields with units of length, so that we can adjoin this modification to the Lagrangians given in Eqs.2.11 and 2.12. Since $L_{BD4ST} \sim l^{-1}$, $L_{E4ST} \sim l^{-2}$ and $A_a \sim l^0$ we see that $\beta$ and $\gamma \sim l^1$ for $L_{BD4ST}$ while $\beta$ and $\gamma \sim l^0$ for $L_{E4ST}$. Thus $\beta$ and $\gamma$ must be constants if $L_{GM}$ is to be adjoined to $L_{E5ST}$, in which case $\gamma$ can be chosen to equal 0, since $\varepsilon^{abcd}F_{ab}F_{cd}$ is a divergence. Hence my choice for a Lagrangian of the Einstein form incorporating five scalar fields, a vector field and a tensor field is

$$L_{E5SVT} = g^{\frac{1}{2}}[R + \sum_{\alpha}\zeta_\alpha(X_\alpha)\varphi_\alpha^{-1}\Box\varphi_\alpha + \sum_{\alpha\leq\beta}\xi_{\alpha\beta}(X_{\alpha\beta})\varphi_\alpha^{-1}\varphi_\beta^{-1} + bF^{ab}F_{ab}] , \qquad \text{Eq.2.16}$$

where b is a constant. Once again, as a preliminary choice for $\zeta_\alpha$ and $\xi_{\alpha\beta}$, we can take the simple forms discussed above.

Figuring out how to adjoin $L_{GM}$ to $L_{BD5ST}$ is a bit more difficult. We know that for this case $\beta$ and $\gamma \sim l^1$, and are functions of $\varphi_0, ..., \varphi_4$. But must all five of these scalar fields appear in $\beta$ and $\gamma$? To determine the answer let's look at how things behave in the solar system. There $\varphi_M$, $\varphi_N$ and $\varphi_P$ vary considerably, and hence the electromagnetic field equation, which will be akin to Eq.2.15, will be quite different from Maxwell's, which is nonsense. So $\varphi_M$, $\varphi_N$ and $\varphi_P$, can not be in either β or γ. Now we expect the dark energy, $\varphi$, to be very constant in the solar system. If we also assume that $\varphi_D$ is fairly constant in the solar system, then β and γ are just functions of $\varphi$ and $\varphi_D$. Thus we see that with these restrictions the electromagnetic field equation, Eq.2.15, will give us Maxwell's equation in the solar system. The ADA tells us that since β and $\gamma \sim l^1$ we must have $\beta = b_0\varphi_0 + b_1\varphi_1$ and $\gamma = d_0\varphi_0 + d_1\varphi_1$, where the b's and d's are numerical constants. Thus physics and mathematics combine to tell us that in this case the dark energy and dark matter must couple to the electromagnetic fields. Hence using Eq.2.11 we find that our modification of $L_{BD4ST}$ to include electromagnetic fields would be

$$L_{BD5SVT} := g^{\frac{1}{2}}[(\sum_\alpha \varphi_\alpha)R + \sum_\alpha \zeta_\alpha(X_\alpha)\Box\varphi_\alpha + \sum_{\alpha \leq \beta}\xi_{\alpha\beta}(X_{\alpha\beta})(\varphi_\alpha\varphi_\beta)^{-\frac{1}{2}} + (b_0\varphi_0+b_1\varphi_1)F^{ab}F_{ab}] +$$
$$+ (d_0\varphi_0+d_1\varphi_1)\varepsilon^{hijk}F_{hi}F_{jk}. \qquad \text{Eq.2.17}$$

As was the case in Eq.2.16, for a first choice, we can take $\zeta_\alpha$ and $\xi_{\alpha\beta}$ to have their simple linear forms. The one thing about the above Lagrangian that you may be uncomfortable with are the terms with $(\varphi_\alpha\varphi_\beta)^{-\frac{1}{2}}$. These terms persist in the Euler-Lagrange equations, and at first seem so out of place when all of the other terms are polynomials (except for $g^{\frac{1}{2}}$). But all of those terms can be removed by simply introducing new scalar fields $\psi_\alpha$ defined by $\psi_\alpha^2 := \varphi_\alpha$. Now that I have rectified that problem, let me remind you that one interesting feature of $L_{BD5SVT}$ is that it requires direct coupling of dark energy and dark matter with the external electromagnetic field, which is something that does not occur with $L_{E5SVT}$. This suggests that one way to distinguish which of the two Lagrangians is more accurate in describing the behavior of the U, is to look for discrepancies in the predictions of Maxwell's equations in regions where we might expect the dark energy or dark matter fields to be varying.

**Section 3: Multiple-Scalar, Vector-Tensor Field Theories Compatible with the ADA**

It is conventionally assumed that the energy-momentum tensor of matter, $T_M^{ab}$, has units of $l^{-2}$ (*see, e.g.,* page 131 in Misner, *et.al.*[16]). Thus if L is the Lagrangian of our theory, and we want $E^{ab}(L) = \kappa T_M^{ab}$, where $\kappa$ is a unitless constant, then we must require $L \sim l^{-2}$, since $g_{ab} \sim l^0$. Now let's assume that our Lagrangian is a function of $n \geq 1$ scalar fields $\{\varphi_\alpha\} = \{\varphi_0, \ldots, \varphi_{n-1}\}$, along with a vector field, $A_a$, a metric tensor, $g_{ab}$, and their derivatives of arbitrary order. It will be assumed that the scalar fields, vector field and tensor field are all unitless. Since L is required to satisfy the ADA we know that for every real number $\lambda > 0$

$$\lambda^{-2}L = L(\varphi_\alpha; \lambda^{-1}\varphi_{\alpha,h}; \lambda^{-2}\varphi_{\alpha,hi}; \ldots; A_a; \lambda^{-1}A_{a,h}; \lambda^{-2}A_{a,hi}; \ldots; g_{ab}; \lambda^{-1}g_{ab,h}; \lambda^{-2}g_{ab,hi}; \ldots).$$

Upon setting $\mu := \lambda^{-1}$, we see that for every $\mu > 0$ the above equation becomes

$$\mu^2 L = L(\varphi_\alpha; \mu\varphi_{\alpha,h}; \mu^2\varphi_{\alpha,hi}; \ldots; A_a; \mu A_{a,h}; \mu^2 A_{a,hi}; \ldots; g_{ab}; \mu g_{ab,h}; \mu^2 g_{ab,hi}; \ldots) \, . \quad \text{Eq.3.1}$$

This equation will be analyzed in the following way. I shall first differentiate it twice with respect to $\mu$. Then a limit will be taken as $\mu \to 0^+$. To evaluate that limit it will be assumed that L is flat space compatible, that is, it is defined and differentiable when either the scalar fields are constant and (or) the vector field is constant, and (or) the metric tensor is flat. Thus when I take the limit as $\mu \to 0^+$ all of the arguments in the derivatives of L having $\mu$ to some positive power can be replaced by 0. It should be noted that the assumption of flat space compatibility puts a severe restriction on the form of L. *E.g.*, it rules out (among other things) coefficients of the form ( $g^{ab}\varphi_{\alpha,a}\varphi_{\beta,b})^{-1}$ from appearing in L. With that said I shall now implement the analysis outlined above. To simplify the form of the ensuing calculations if $T^{\cdots}$ is any quantity I shall set

$$T^{\cdots;\alpha,hi\cdots} := \frac{\partial T^{\cdots}}{\partial \varphi_{\alpha,hi\cdots}} \, , \quad T^{\cdots;a,hi\cdots} := \frac{\partial T^{\cdots}}{\partial A_{a,hi\cdots}} \quad \text{and} \quad T^{\cdots;ab,hi\cdots} := \frac{\partial T^{\cdots}}{\partial g_{ab,hi\cdots}} \, .$$

I shall also let, say, $L^{;\alpha,hi\cdots}(S)$, denote the partial derivative of L with respect to $\varphi_{\alpha,hi\cdots}$ evaluated for all the "stuff" appearing as arguments on the right-hand side of Eq.3.1. So if we now differentiate Eq.3.1 with respect to $\mu$ we find that

$$2\mu L = \sum_\alpha L^{;\alpha,h}(S)\varphi_{\alpha,h} + 2\mu\sum_\alpha L^{;\alpha,hi}(S)\varphi_{\alpha,hi} + \mu^2[\text{Junk from } \varphi_{\alpha,hi\cdots} \text{ derivatives}] + L^{;a,h}(S)A_{a,h}$$

$$+ \, 2\mu L^{;a,hi}(S)A_{a,hi} + \mu^2[\text{Junk from } A_{a,hi\cdots} \text{ derivatives}] + L^{;ab,h}(S)g_{ab,h} +$$

$$+ \, 2\mu L^{;ab,hi}(S)g_{ab,hi} + \mu^2[\text{Junk from } g_{ab,hi\cdots} \text{ derivatives}]. \qquad \text{Eq.3.2}$$

Upon differentiating Eq.3.2 with respect to µ, we see that

$$2L = \sum_{\alpha,\beta} L^{;\alpha,h;\beta,i}(S)\varphi_{\alpha,h}\varphi_{\beta,i} + \sum_{\alpha} L^{;\alpha,h;a,i}(S)\varphi_{\alpha,h}A_{a,i} + \sum_{\alpha} L^{;\alpha,h;ab,i}(S)\varphi_{\alpha,h}g_{ab,i} + 2\sum_{\alpha} L^{;\alpha,hi}(S)\varphi_{\alpha,hi} +$$

$$+ \sum_{\alpha} L^{;a,h;\alpha,i}(S)\varphi_{\alpha,i}A_{a,h} + L^{;a,h;b,i}(S)A_{a,h}A_{b,i} + L^{;a,h;bc,i}(S)A_{a,h}g_{bc,i} + 2L^{;a,hi}(S)A_{a,hi} +$$

$$+ \sum_{\alpha} L^{;ab,h;\alpha,i}(S)g_{ab,h}\varphi_{\alpha,i} + L^{;ab,h;c,i}(S)g_{ab,h}A_{c,i} + L^{;ab,h;cd,i}(S)g_{ab,h}g_{cd,i} + 2L^{;ab,hi}(S)g_{ab,hi} +$$

$$+ \mu[\text{Junk}] . \qquad \text{Eq.3.3}$$

When we take the limit of Eq.3.3 as $\mu \to 0^+$, recalling that L is assumed to be flat space compatible, we discover that

$$L = \sum_{\alpha \leq \beta} \Lambda^{\alpha\beta hi}\varphi_{\alpha,h}\varphi_{\beta,i} + \sum_{\alpha} \Lambda^{\alpha hai}\varphi_{\alpha,h}A_{a,i} + \sum_{\alpha} \Lambda^{\alpha habi}\varphi_{\alpha,h}g_{ab,i} + \sum_{\alpha} \Lambda^{\alpha hi}\varphi_{\alpha,hi} + \Lambda_1^{ahbi}A_{a,h}A_{b,i} +$$

$$+ \Lambda^{ahbci}A_{a,h}g_{bc,i} + \Lambda^{ahi}A_{a,hi} + \Lambda^{abhcdi}g_{ab,h}g_{cd,i} + \Lambda_2^{abhi}g_{ab,hi} , \qquad \text{Eq.3.4}$$

where the $\Lambda$'s are obtained from the partial derivatives of L, and are concomitants of only the $\varphi_\alpha$'s, $A_a$ and $g_{ab}$. The $\Lambda$'s have the obvious symmetries, inherited from what they are summing into; *e.g.,* $\Lambda^{ahbci}$ is symmetric in b and c. In addition $\Lambda^{\alpha hi}$, $\Lambda^{ahi}$ and $\Lambda_2^{abhi}$ are tensorial concomitants. This follows from the fact that when you differentiate a second-order, multiple-scalar-vector-tensor concomitant, with respect to either

$\varphi_{\alpha,hi}$ , $A_{a,hi}$ or $g_{ab,hi}$, you obtain tensorial quantities.

To simplify the form of Eq.3.4 we require the following

**Lemma 1 (Thomas's Replacement Theorem for Multiple-Scalar-Vector Tensor Concomitants):** If τ is a second-order tensorial concomitant which locally has the

form

$$\tau^{\cdots}_{\cdots} = \tau^{\cdots}_{\cdots}(\varphi_\alpha; \varphi_{\alpha,h}; \varphi_{\alpha,hi}; A_a; A_{a,h}; A_{a,hi}; g_{ab}; g_{ab,h}; g_{ab,hi}),$$

then the value of τ's components are unaffected if their arguments are replaced as indicated below:

$$\tau^{\cdots}_{\cdots} = \tau^{\cdots}_{\cdots}(\varphi_\alpha; \varphi_{\alpha h}; \varphi_{\alpha hi}; A_a; A_{a|h}; A_{a|(hi)} + \tfrac{1}{6}A_m(R_h{}^m{}_{ai} + R_i{}^m{}_{ah}); g_{ab}; 0; \tfrac{1}{3}(R_{ahib}+R_{aihb}))$$

where for the scalar fields $\varphi_\alpha$ we let $\varphi_{\alpha h \ldots i} := \varphi_{\alpha | h \ldots i}$.

**Proof:** In [17] and [18] I essentially explain how Thomas's Replacement Theorem presented in [19] gives rise to the result presented above.■

Due to Thomas's Replacement Theorem Eq.3.4 implies that

$$L = \sum_{\alpha \leq \beta} \Lambda^{\alpha\beta hi}\varphi_{\alpha h}\varphi_{\beta i} + \sum_\alpha \Lambda^{\alpha hai}\varphi_{\alpha h}A_{a|i} + \sum_\alpha \Lambda^{\alpha hi}\varphi_{\alpha hi} + \Lambda_1^{ahbi}A_{a|h}A_{b|i} + \Lambda^{ahi}(A_{a|hi} + \tfrac{1}{3}A_m R_h{}^m{}_{ai})$$
$$+ \tfrac{2}{3}\Lambda_2^{abhi}R_{ahib}. \qquad \text{Eq.3.5}$$

Above I mentioned that $\Lambda^{\alpha hi}$, $\Lambda^{ahi}$ and $\Lambda_2^{abhi}$ are tensorial concomitants, so if we take the terms in Eq.3.5 involving these quantities to the left-hand side of Eq.3.5 what remains on the right-hand side must be a first-order tensorial concomitant of $\varphi_\alpha$, $A_a$ and $g_{ab}$. Consequently the derivatives of what remains on the right-hand side of Eq.3.5 when we differentiate with respect to $\varphi_{\alpha,i}$ and $A_{a,i}$ must be tensorial. In this way we can show that $\Lambda^{\alpha\beta hi}$, $\Lambda^{\alpha hai}$ and $\Lambda_1^{ahbi}$ must be tensorial concomitants of $\varphi_\alpha$, $A_a$ and $g_{ab}$. In [17] I explain in some detail how one goes about constructing tensorial concomitants of this form. This is also done in great detail in Horndeski [20].

Employing the theory developed in these cited works it is relatively easy to prove that:

$$\Lambda^{\alpha\beta hi} = g^{\frac{1}{2}}(\Lambda_1{}^{\alpha\beta}g^{hi} + \Lambda_2{}^{\alpha\beta}A^hA^i), \qquad \text{Eq.3.6}$$

$$\Lambda^{\alpha hi} = g^{\frac{1}{2}}(\Lambda_1{}^{\alpha}g^{hi} + \Lambda_2{}^{\alpha}A^h A^i), \qquad \text{Eq.3.7}$$

$$\Lambda^{\alpha hai} = g^{\frac{1}{2}}(\Xi_1{}^{\alpha}A^hg^{ai} + \Xi_2{}^{\alpha}A^ag^{ih} + \Xi_3{}^{\alpha}A^ig^{ha} + \Xi_4{}^{\alpha}A^hA^aA^i) + \Xi_5{}^{\alpha}\varepsilon^{haib}A_b, \qquad \text{Eq.3.8}$$

and

$$\Lambda^{ahi} = g^{\frac{1}{2}}(\Lambda_1 A^ag^{hi} + \Lambda_2 A^h g^{ia} + \Lambda_2 A^ig^{ah}), \qquad \text{Eq.3.9}$$

where the $\Lambda$'s and $\Xi$'s are scalar concomitants of $\varphi_\alpha$ and $Z := g_{ab}A^a A^b$. We still have two (4,0) tensor densities to determine, $\Lambda_1{}^{abhi}$ and $\Lambda_2{}^{abhi}$. The general form of a (4,0) tensor density concomitant, $\Omega^{abhi}$, of $\varphi_\alpha$, $A_a$ and $g_{ab}$ is given by

$$\Omega^{abhi} = g^{\frac{1}{2}}(\Omega_1 g^{ab}g^{hi} + \Omega_2 g^{ah}g^{ib} + \Omega_3 g^{ai}g^{bh} + \Omega_4 g^{ab}A^hA^i + \Omega_5 g^{ah}A^iA^b + \Omega_6 g^{ai}A^bA^h +$$

$$+ \Omega_7 g^{bh}A^aA^i + \Omega_8 g^{bi}A^aA^h + \Omega_9 g^{hi}A^aA^b + \Omega_{10}A^aA^bA^hA^i) + \Omega_{11}\varepsilon^{abhi}, \qquad \text{Eq.3.10}$$

where the $\Omega$'s are scalar concomitants of $\varphi_\alpha$ and Z. Now $\Lambda_1{}^{abhi} = \Lambda_1{}^{hiab}$, and $\Lambda_2{}^{abhi} = \Lambda_2{}^{bahi} = \Lambda_2{}^{abih}$. So we can use the expression presented in Eq.3.10 to produce $\Lambda_1{}^{abhi}$ and $\Lambda_2{}^{abhi}$ with the proper symmetries. Using these expressions, along with those given in Eqs.3.6-3.9, we find, after a great deal of simplification, that if we let $Y_\alpha := g_{ab}\varphi_\alpha{}^aA^b$, then

$$L = g^{\frac{1}{2}}[\sum_{\alpha \leq \beta}\Lambda_1{}^{\alpha\beta}Y_\alpha Y_\beta + \sum_\alpha \Lambda_2{}^{\alpha\beta}X_{\alpha\beta} + (\sum_\alpha \Lambda_1{}^{\alpha}A^hA^i\varphi_{\alpha h} + \sum_\alpha \Lambda_2{}^{\alpha}\varphi_\alpha{}^i)_{|i} + \sum_\alpha \Xi_1{}^{\alpha}\varphi_\alpha{}^hA^iA_{h|i} +$$

$$+ \sum_\alpha \Xi_2{}^{\alpha}\varphi_{\alpha h}Z^{|h} + \sum_\alpha \Xi_3{}^{\alpha}Y_\alpha A^a{}_{|a} + \sum_\alpha \Xi_4{}^{\alpha}Y_\alpha Z_{|i}A^i + \Omega_1 A^h{}_{|h}A^i{}_{|i} + \Omega_2 A^{h|i}A_{h|i}$$

$$+ \Omega_3 A^{h|i}A_{i|h} + \Omega_4 A^hA^iA_{h|i}A^j{}_{|j} + \Omega_5 A^hA^iA^j{}_{|h}A_{j|i} + \Omega_6 A^hA^iA^j{}_{|h}A_{i|j} + \Omega_7 A^hA^iA_h{}^{|j}A_{i|j} +$$

$$+ \Omega_8 A^h A^i A^j A^k A_{h|i} A_{j|k} + (\Lambda_1 A^h A_h{}^{|i} + \Lambda_2 A^h A^{i|h})_{|i} + \Lambda_3 A^h A^i R_{hi} + \Lambda_4 R] +$$

$$+ \sum_\alpha \Xi_5{}^\alpha \varepsilon^{hbai} \varphi_{\alpha h} A_b A_{a|i} + \Omega_9 \varepsilon^{hijk} A_{h|i} A_{j|k} ,\qquad \text{Eq.3.11}$$

where the $\Lambda$'s, $\Xi$'s and $\Omega$'s appearing in Eq.3.11 are unitless scalar concomitants of the $\varphi_\alpha$'s and Z, and are not exactly the same ones that appear in Eqs.3.6-3.9, since terms have been combined and numerical factors have been absorbed. The important thing to note about Eq.3.11 is that each term has (of course) units of $l^{-2}$, is (up to a divergence) first-order in $\varphi_\alpha$ and $A_a$, and at most second-order in $g_{ab}$; and, lastly, yields second-order Euler-Lagrange tensors. Thus this Lagrangian will generate field equations that are free of Ostrogradsky [21] instabilities.

The above work is summarized in the following

**Theorem:** In an orientable, four-dimensional pseudo-Riemannian space, suppose that L is a Lagrangian which is a concomitant of n≥1 scalar fields, $\varphi_\alpha$, a vector field, $A_a$, and a pseudo-Riemannian metric tensor, $g_{ab}$, along with their derivatives of arbitrary, but finite, order. If L~$l^{-2}$, satisfies the Axiom of Dimensional Analysis, and is flat space compatible, then L must have the form of the Lagrangian presented in Eq.3.11. Consequently L must be at most of second-order, and yields Euler-Lagrange tensor densities which are at most of second-order.∎

It should be noted that if we only have one scalar field and no vector field, then the above result agrees with what Aldersley presents in [13]. In addition, if there is only one scalar field, then the Lagrangian presented in Eq.3.11 is among the class of scalar-vector-tensor Lagrangians presented by Heisenberg in [22] that yield second-order field equations.

A multiple-scalar-vector-tensor Lagrangian will be said to be gauge invariant if the Lagrangian is invariant when the vector potential $A_a$ (and its derivatives) are replaced throughout the Lagrangian by $A_a + \psi_{,a}$ (and its derivatives), where $\psi$ is an arbitrary scalar field. When a Lagrangian is gauge invariant, so are its associated Euler-Lagrange tensor densities, and it is well-known that this implies that the Euler-Lagrange equations are compatible with conservation of charge (*see, e.g.,* Horndeski [23]). In [23] I investigate gauge invariant Lagrangians, and a result from that paper that we shall need is presented in

**Lemma 2:** If L satisfies the assumptions of the Theorem and is gauge invariant, then L must be of second-order and

$$L(\varphi_\alpha; \varphi_{\alpha,h}; \varphi_{\alpha,hi}; A_a; A_{a,h}; A_{a,hi}; g_{ab}; g_{ab,h}; g_{ab,hi}) =$$
$$= L(\varphi_\alpha; \varphi_{\alpha,h}; \varphi_{\alpha,hi}; 0; \tfrac{1}{2}F_{ab}; \tfrac{2}{3}F_{a(h,i)}; g_{ab}; g_{ab,h}; g_{ab,hi})$$

where $F_{ab} := A_{a,b} - A_{b,a}$. ∎

As an immediate consequence of **Lemma 2**, and the **Theorem**, we have the following

**Corollary:** In an orientable, four-dimensional space, suppose that L is a Lagrangian which is a concomitant of n $\geq 1$ scalar fields, $\varphi_\alpha$, a vector field, $A_a$, and a pseudo-Riemannian metric tensor, $g_{ab}$, along with their derivatives of arbitrary, but finite, order. If $L \sim l^{-2}$, satisfies the ADA, is flat space compatible and invariant under gauge transformations, then L must be given by

$$L = g^{\frac{1}{2}}[\sum_{\alpha \leq \beta} \lambda^{\alpha\beta} X_{\alpha\beta} + \omega_1 F^{ab} F_{ab} + \lambda R] + \omega_2 \varepsilon^{hijk} F_{hi} F_{jk} \,, \qquad \text{Eq.3.13}$$

where $\lambda$, $\omega_1$, $\omega_2$ and $\lambda^{\alpha\beta}$ are arbitrary scalar concomitants of the $\varphi_\alpha$'s. ∎

Once again, for the case where there is one scalar field, the Corollary is in keeping with the results presented by Heisenberg in [22].

In view of the simplicity of the Lagrangian presented in Eq.3.13, and its similarity to the Lagrangians used in Einstein's theory, it seems fairly plausible that $c_g$ will equal 1 for the field equations it generates. I even suspect that $c_g = 1$, for the field equations associated with the Lagrangian presented in Eq.3.11, but I have no proof of that proposition.

**Section 4: Concluding Remarks**

The work of the previous section shows that it is fairly straightforward to use dimensional analysis to generate Lagrangians which satisfy the ADA, provided all of the field varilables are unitless, and you demand flat space compatibility. You simply choose what you want the units of L to be, and then proceed as was done in **Section**

**3**. Problems only arise when some of the fields under consideration have units of $l^\mu$ where $\mu>0$. In **Section 2** I showed how that situation can be dealt with for certain scalar-tensor field theories, and discussed possible extensions to multiple-scalar-vector-tensor field theories.

I have recently come to believe that the effects ascribed to dark energy and dark matter can be represented by a field theory involving two scalar fields in conjunction with the metric tensor. The problem is how to incorporate these fields into a bi-scalar-tensor field theory. Suppose we let $\psi_0$ and $\psi_1$ denote unitless scalar fields intended to account for the effects of dark energy and dark matter. If $L_{2ST}$ is a bi-scalar-tensor Lagrangian for the associated field theory, with units of $(length)^{-2}$ satisfying the ADA, and the flat space compatibility condition, then due to the **Theorem**, $L_{2ST}$ must be given by

$$L_{2ST} = g^{½}[\Xi_{00}g^{ab}\psi_{0,a}\psi_{0,b} + \Xi_{01}g^{ab}\psi_{0,a}\psi_{1,b} + \Xi_{11}g^{ab}\psi_{1,a}\psi_{1,b} + \Xi R] , \qquad \text{Eq.4.1}$$

where the $\Xi$'s are scalar functions of $\psi_0$ and $\psi_1$. Frankly, the Lagrangian $L_{2ST}$, is not that exciting and fairly ordinary. It is pretty much the obvious Lagrangian to employ in the conventional approach to General Relativity, and I believe that we require something different.

From what I have done so far in this paper, you know that I am trying very hard to not introduce any new dimensioned constants. But now is the time to do so.

Let λ be a constant with units of length. I define two scalar fields $\varphi_0$ and $\varphi_1$ by $\varphi_0 := \lambda \psi_0$ and $\varphi_1 := \lambda \psi_1$, where $\psi_0 \sim l^0$ represents dark energy effects, and $\psi_1 \sim l^0$ represents dark matter effects. The theory developed in **Section 2** ( in particular Eqs.2.11 and 2.12) involving five scalar fields with units of length, and a unitless vector field, can now be called upon to provide us with a suitable Lagrangian to represent a cosmological theory describing the combination of dark energy and dark matter with ordinary matter, neutrinos, photons and electromagnetic fields. In particuar I am very fond of two Lagrangians for which simple choices of the functions $\zeta_\alpha$ and $\xi_{\alpha\beta}$ are made. These Lagrangians are

$$\mathcal{L}_{E5SVT} := g^{\frac{1}{2}}[R + k_0 X_0\, \varphi_0^{-1}\Box\varphi_0 + k_1 X_1\, \varphi_1^{-1}\Box\varphi_1 + \sum_{\alpha \leq \beta} k_{\alpha\beta} X_{\alpha\beta}\, \varphi_\alpha^{-1}\varphi_\beta^{-1} +$$
$$+ bF^{ab} F_{ab}]\,, \qquad \text{Eq.4.2}$$

and

$$\mathcal{L}_{BD5ST} := g^{\frac{1}{2}}[(\varphi_0+\varphi_1)R + k_0 X_0 \Box\varphi_0 + k_1 X_1 \Box\varphi_1 + \sum_{\alpha \leq \beta} k_{\alpha\beta} X_{\alpha\beta} (\varphi_\alpha \varphi_\beta)^{-\frac{1}{2}} +$$
$$+ (b_0 \varphi_0 + b_1 \varphi_1) F^{ab} F_{ab}] + (d_0 \varphi_0 + d_1 \varphi_1)\varepsilon^{hijk} F_{hi}\, F_{jk}\,, \qquad \text{Eq.4.3}$$

where the k's, b's and d's are constants that need to be determined. Note that the Lagrangians given in Eqs.4.2 and 4.3 are not the same as those presented in Eqs.2.11 and 2.12, since I only let the sums over α go from 0 to 1, while the sums over α≤β have their full range of 0 to 4. I did this because I thought that it would be more in keeping with $\varphi_0$ and $\varphi_1$ representing gravitational phenomenon, and it also mirrors their appearance as coefficients of $F^{ab}F_{ab}$ and $\varepsilon^{hijk}F_{hi}\, F_{jk}$.

In conclusion I need to remark upon the one assumption that is common to most field theories; *viz*., that the equations of the field theory be derivable from a principle of least action. Nietzsche used to make fun of us for doing that by saying something like "the principle of least action is also the principle of greatest stupidity." The reason for his callous remark is fairly obvious, since we are all waiting for the proof that the equations governing the natural world must come from a variational principle, which has been our pole star in the quest to understand the U.

**Acknowledgement**

I would like to thank Dr. M. Zumalaćarregui for discussions on the topic of field theories involving multiple scalar fields.